\documentclass{article}
\usepackage[utf8]{inputenc}
\usepackage[yyyymmdd]{datetime}
\usepackage{url}
\usepackage{tabularx}
\usepackage{tikz}
\usepackage{wasysym}
\usepackage{flowchart}
\usepackage{breakurl}
\usetikzlibrary{shapes,arrows.meta,chains}

\title{Token-Based Payment Systems}
\author{Geoffrey Goodell}
\date{\small \textit{This Version: \today}}

\newcommand{\cz}[1]{\textit{\textbf{#1}}}

\newcolumntype{L}[1]{>{\raggedright\arraybackslash}p{#1}}
\newcolumntype{C}[1]{>{\centering\arraybackslash}p{#1}}
\newcolumntype{R}[1]{>{\raggedleft\arraybackslash}p{#1}}

\newcommand{\ts}{
    \tikzset{>={Latex[width=3mm,length=3mm]}}
    \tikzstyle{line} = [draw, ->, >=latex, ultra thick]
    \tikzstyle{box} = [
      draw,
      rectangle,
      thick,
      align=center,
      text width=3.2cm,
      text centered,
      minimum height=2.5em
    ]
    \tikzstyle{noshape} = [align=center, text centered]
}

\begin{document}

\maketitle

\begin{abstract}

In this article, we consider the roles of tokens and distributed ledgers in
digital payment systems.  We present a brief taxonomy of digital payment
systems that use tokens, and we address the different models for how
distributed ledger technology can support digital payment systems in general.
We offer guidance on the salient features of digital payment systems, which we
comprehend in terms of consumer privacy, token issuance, and accountability for
system operators.

\end{abstract}

\section{Introduction}

Modern digital payments make use of tokens and distributed ledgers in various
ways.  Digital tokens underpin much of decentralised finance in the form of
cryptocurrency, and they can also be used to create central bank digital
currency and stablecoins.  Although tokens and distributed ledgers are often
found together, tokens can be used without distributed ledgers, and distributed
ledgers can be used without tokens.

The use of digital tokens as a means of payment is not new.  The core problem
underpinning the tussle surrounding digital payment mechanisms is the absence
of a perfect substitute for physical cash.  Cash is a \textit{sui generis} kind
of asset, uniquely functioning as both a physical possession and a claim on the
central bank.  In particular, account-based money, such as that which is
offered by banks and closed-loop payment systems, are intrinsically linked to
the identities of their users and subject to the performance of their
fiduciaries, whereas token-based money can be used as bearer instruments.  From
the perspective of consumers, using accounts to make payments carries a set of
intrinsic costs and risks associated with custodial relationships, including
unwelcome surveillance and gatekeeping enabled by the power relationship
between the account-holder and the fiduciary.  Such costs and risks have been
recognised for over five decades~\cite{armer1968,armer1975} and motivated the
development of token-based digital currency~\cite{chaum1982}.

A distributed ledger is a ``ledger that is shared across a set of DLT nodes and
synchronized between the DLT nodes using a consensus
mechanism''~\cite{iso22739}.  Distributed ledger technology (DLT) allows a set
of participating peers to externalise their commitments, thus rendering
themselves mutually accountable by establishing a shared version of the history
of transactions.  Cryptocurrencies are often implemented as decentralised
systems with permissionless DLT systems at their core, although DLT can be used
in other ways to support electronic payments.

This report considers the different manifestations of token-based payment
systems and the use of distributed ledgers to support payment systems in
general.  We provide a taxonomy of token-based payment systems according to
their technical characteristics, and we characterise the various ways in which
DLT can enable digital payments.

\section{Taxonomy of token-based digital payment systems}

Token-based digital currency has many manifestations, including cryptocurrency,
private digital currencies, and central bank digital currency.  The Bank for
International Settlements has characterised different forms of money in terms
of \textit{administrative} dimensions of digital currency, such as whether it
is publicly available, whether it can be transacted in a peer-to-peer way, and
whether it is issued by a central bank~\cite{bis2017}.  However, it is also
possible to consider the following \textit{technical} dimensions of digital
currency:

\begin{itemize}

\item whether the ledger is centralised,

\item whether transactions are private by design, and

\item whether the tokens are intrinsic to the consensus mechanism.

\end{itemize}

The three technical dimensions described above are useful for understanding the
implicit accountability relationships among users and system operators, which
in turn provide insight into the use cases for which particular token-based
payment systems might be appropriate.  As with all payment systems, different
token-based payment methods carry different sets of associated costs and risks,
and therefore might be more or less suitable in different contexts.  For
example, purchasing securities and purchasing consumer goods might necessitate
different payment mechanisms.

\begin{figure}[ht]
\begin{center}
\hspace{-0.8em}\scalebox{1.2}{
\begin{tikzpicture}[>=latex, node distance=3cm, font={\sf \small}, auto]\ts
\tikzset{>={Latex[width=4mm,length=4mm]}}
\node (box1) at (-2,2) [box, text width=10.7em, text height=10.7em] {};
\node (box2) at (2,2) [box, text width=10.7em, text height=10.7em] {};
\node (box3) at (-2,-2) [box, text width=10.7em, text height=10.7em] {};
\node (box4) at (2,-2) [box, text width=10.7em, text height=10.7em] {};
\node (n1) at (-2,4.5) {centralised};
\node (n2) at (2,4.5) {decentralised};
\node[align=center] (t1) at (-2,2) {
    electronic vouchers\\
    \\
    (e.g. store credits)
};
\node[align=center] (t2) at (2,2) {
    \vspace{0.5em}\\
    most UTXO\\
    cryptocurrency\\
    \\
    (e.g. Bitcoin, Litecoin)\\
};
\node[align=center] (t2) at (-2,-2) {
    DigiCash, e-gold\\
    \\
    Chaumian CBDC
};
\node[align=center] (t2) at (2,-2) {
    Monero, Zcash
};
\node[rotate=90] (n3) at (-4.5,2) {transparent};
\node[rotate=90] (n4) at (-4.5,-2) {private};
\end{tikzpicture}}
\end{center}

\caption{\cz{Token-based digital payment systems whose ledgers track the state
of individual assets.}}

\label{f:ledger-track}
\end{figure}

There are two principal methods to track tokens using a ledger:
\textit{endogenous}, wherein the ledger maintains the state of each asset, and
\textit{oblivious}, wherein assets maintain their own state.  Endogenous
tracking generally involves representing the status of the tokens themselves or
the results of individual transactions directly on the ledger.  With oblivious
tracking, a \textit{proof of provenance} is maintained and transferred along
with the assets, and the ledger is used only for verification.

With endogenous tracking, tokens are generally intrinsic to the operation of
the ledger and cannot be moved from one ledger to another.
Figure~\ref{f:ledger-track} offers a classification of token-based digital
payment systems that use endogenous tracking.  Such systems can be:

\begin{itemize}

\item\cz{centralised and transparent}, wherein tokens are created, maintained, and
accepted by a single authority, as is the case with electronic vouchers;

\item\cz{centralised and private}, wherein tokens are issued in a manner in which
they cannot be individually recognised when they are spent, using blind
signatures or other privacy-enhancing technology;

\item\cz{decentralised and transparent}, wherein tokens are created and transacted
within a distributed ledger, using a mechanism that tracks the state of the
output of each transaction and the flow of assets can be monitored by observers
of the ledger; or

\item\cz{decentralised and private}, wherein tokens are created and transacted
within a distributed ledger, using a mechanism that tracks the state of the
output of each transaction but the flow of assets cannot be monitored by
observers of the ledger.

\end{itemize}

Payment systems with endogenous token tracking are also called \textit{unspent
transaction output} (UTXO) systems because of the method they use to ensure
that every token can be spent exactly once.  During each transaction of a
particular value, a set of new tokens with that specific value is created, and
the ledger incorporates a record of the tokens involved in specific
transactions to ensure that tokens are not spent twice.  The protocol developed
by David Chaum in 1982~\cite{chaum1982} and commercialised as
DigiCash~\cite{digicash} is an example of a centralised UTXO system with a
privacy-preserving ledger that allows untraceable payments.  Chaum's more
recent work with the Swiss National Bank~\cite{chaum2021} follows the same
principle.  Bitcoin~\cite{nakamoto} is an example of a decentralised UTXO
system with a transparent ledger.  For the architects of Bitcoin, the purpose
of decentralisation is to provide immutability, ensuring that system operators
would not be able to reverse a transaction~\cite{nakamoto}.  Unlike DigiCash
tokens, however, Bitcoin tokens are neither private nor truly fungible, since
the use of specific tokens can be linked to their individual creation, thus
allowing the history of assets to be traced~\cite{anderson2018}.
Privacy-oriented UTXO cryptocurrencies, notably
Monero~\cite{vansaberhagen2013}, address this limitation by combining both
decentralised transactions and privacy by design.

\begin{figure}[ht]
\begin{center}
\hspace{-0.8em}\scalebox{1.2}{
\begin{tikzpicture}[>=latex, node distance=3cm, font={\sf \small}, auto]\ts
\tikzset{>={Latex[width=4mm,length=4mm]}}
\node (box1) at (-2,2) [box, text width=10.7em, text height=10.7em] {};
\node (box2) at (2,2) [box, text width=10.7em, text height=10.7em] {};
\node (box3) at (-2,-2) [box, text width=10.7em, text height=10.7em] {};
\node (box4) at (2,-2) [box, text width=10.7em, text height=10.7em] {};
\node (n1) at (-2,4.5) {centralised};
\node (n2) at (2,4.5) {decentralised};
\node[align=center] (t1) at (-2,2) {
    standard USO assets
};
\node[align=center] (t2) at (2,2) {
    USO assets with\\
    mitigation against\\
    equivocation
};
\node[align=center] (t2) at (-2,-2) {
    fungible USO assets
};
\node[align=center] (t2) at (2,-2) {
    fungible USO assets\\
    with mitigation\\
    against equivocation
};
\node[rotate=90] (n3) at (-4.5,2) {transparent};
\node[rotate=90] (n4) at (-4.5,-2) {private};
\end{tikzpicture}}
\end{center}

\caption{\cz{Token-based digital payment systems whose assets track their own
state.}}

\label{f:oblivious}
\end{figure}

Payment systems with oblivious tracking do not use the ledger to track
individual tokens.  Instead, the assets contain their own histories in the form
of a proof of provenance, and the ledger is used to assess the validity of
assets by verifying that the transactions referenced in the proof of provenance
had taken place and that no other transactions had taken place.  Authenticity
of an asset can be assessed by verifying a signature contained within the asset
at the time of its initial creation and satisfaction of rules that apply to
successive transactions.  Creating or transacting an asset requires posting the
hash value of the transaction to the ledger.  Because it is not necessary for
the operator of the ledger to know anything about what is being transacted, we
can say that the operator is \textit{oblivious} to the transaction, and the
assets themselves can be described as unforgeable, stateful and oblivious, or
\textit{USO assets}~\cite{goodell2022}.  Token issuance is separate from ledger
consensus, and because assets track their own state, double-spending is avoided
without the need for the issuer or the ledger to maintain a database of the
individual assets.  Figure~\ref{f:oblivious} illustrates how payment systems
with oblivious tracking can be organised into categories:

\begin{itemize}

\item\cz{centralised and transparent}, wherein each asset contains its complete
history including the details of its creation, as is the case, for example,
with TODA assets~\cite{todapop};

\item\cz{centralised and private}, wherein asset issuance uses privacy-enhancing
technology such as blind signatures~\cite{ietf-blind} to decouple an asset from
its initial creation;

\item\cz{decentralised and transparent}, wherein the system is operated by a
set of peers that externalise their commitments by periodically publishing the
hash value of their accumulated set of transactions to the distributed ledger,
but for which the assets themselves are transparent; and

\item\cz{decentralised and private}, similarly to the case for decentralised
and transparent systems for oblivious tracking, except that details of the
issuance of a particular asset are obscured using privacy-enhancing
technology~\cite{goodell2022}.

\end{itemize}

In principle, payment systems with oblivious tracking depend upon the system
operator not \textit{equivocating}, or signing proofs of provenance with
different versions of history, although this can be mitigated through the use
of DLT.  The effectiveness of the mitigation is determined in part by the
independence of the participants in the distributed ledger.  Privacy-enhancing
technology serves the same purpose as for payment systems with endogenous token
tracking, to separate token issuance from the transaction in which it is spent.
Observe that because commitments to the distributed ledger are periodic and
only involve hash values over an accumulated set of transactions, the size of
the ledger does not grow with the number of transactions.  In exchange for this
efficiency, users of the assets bear the burden of verifying the authenticity
of assets and the validity of their proofs of provenance.

\section{The use of distributed ledger technology in digital payments}

Ledgers are designed to be immutable~\cite{iso22739}.  Distributed ledgers are
designed to achieve immutability by soliciting and recording consensus among a
set of peers about the specific history of transactions.  As a result, a record
can be considered authentic without the requirement to trust any particular
operator of the ledger system to behave as promised.  In principle, with a set
of independent peers, transactions that had previously been validated will not
be repudiable, and claimed transactions that had not previously been validated
by consensus will be recognised as invalid.  Because commitments by the system
operators are externalised, they can be independently verified.

DLT can support both payment systems that maintain balances representing the
assets held by identified users (account-based systems) and payment systems
that allow assets in the form of tokens to be transferred between transacting
parties (token-based systems).  DLT can support these systems in several ways:

\begin{itemize}

\item\cz{Recording transactions between accounts.}  For account-based
payment systems, DLT can be used to provide a record of transactions,
potentially reducing the costs and risks of pairwise account reconciliation
procedures.  In particular, DLT can be used in the following ways:

\begin{enumerate}

\item \textit{To provide evidence that value has been transferred between
externally managed accounts.}  If accounts are managed independently by
fiduciaries, then a distributed ledger can be used to record commitments by
those fiduciaries that a transaction between those accounts has occurred.  By
providing a way for the fiduciaries to externalise their commitments, the DLT
system allows claims about the authenticity of a transaction to be
independently verified.

\item \textit{To implement transfers of value between externally managed
accounts.}  If accounts are managed independently by fiduciaries that agree to
use a distributed ledger as the official record of their transactions, then the
distributed ledger can record the size of the transaction and identifiers
representing the accounts of the two counterparties.  By providing a way for
the fiduciaries to externalise their commitments, the DLT system allows claims
about the authenticity of a transaction to be independently verified.

\item \textit{To implement transfers of value between internally managed
accounts.}  A DLT system can be used to directly manage accounts on behalf of
its users.  The use of balances rather than tokens forms the basis of some
cryptocurrency systems, notably Ethereum~\cite{buterin2014}, which uses a DLT
system to record messages representing the state transitions that result from
executing the transaction.  The immutability of the ledger ensures that state
transitions are non-repudiable, allowing claims about the authenticity of a
transaction or its effect on the balances of transacting parties to be
independently verified.

\end{enumerate}

\item\cz{Managing tokens in payment systems with endogenous token
tracking.}  Payment systems with endogenous token tracking are generally
designed so that tokens can be spent at most once.  During a transaction, old
tokens are retired, and new tokens are created in their place.  Transactions
are recorded on the ledger, referencing an address that specifies the
destination of an asset during a transfer.  The ledger also records an
identifier specifying the token that has been spent, either along with the
transaction or as part of the transaction, for the purpose of ensuring that the
token will not be spent twice.  Depending upon the specific architecture,
transactions can contain the identifiers of new tokens that are created
(``minted'') during a transaction.  However, privacy-preserving UTXO systems,
such as Monero, use privacy-enhancing technology, such as blind signatures or
zero-knowledge proofs, to avoid revealing information that links the
transaction in which a token was created to the transaction in which a token is
spent.  In all cases, the immutability of the ledger allows claims about the
validity of a token to be independently verified at the time that it is spent.

\item\cz{Memorialising commitments in payment systems with oblivious token
tracking.}  In payment systems with oblivious token
tracking~\cite{todapop,goodell2022}, each token tracks its own state, which is
updated during a transaction.  The state of an asset generally includes an
encumbrance that prevents its state from being updated by anyone other than the
user that controls it, and control is generally demonstrated by possession of a
private key.  Assets can be created or transferred between users via state
updates, whose hash values are delivered to a third-party system operator, who
uses the hash values to incorporate the updates into recorded history in a way
that can be proven by the users.  Because the system operator receives only
hash values, it generally does not learn the details of assets being
transferred or even both counterparties to the transaction.  The system
operator furnishes a proof of provenance demonstrating that a specific sequence
of updates to the asset has taken place.  For this approach to work,
counterparties to the transaction must trust that the system operator does not
\textit{equivocate} by presenting different versions of history.

DLT can be used to prevent equivocation by providing a way for system operators
to externalise their commitments among a set of peers.  Each system operator
commits to a specific version of history by periodically publishing to a
distributed ledger a hash value that represents the accumulated set of
transactions that it has received over some period of time.  Because every
proof of provenance received from a system operator can be cross-referenced
against the hash values in the distributed ledger, the claims intrinsic to the
proofs of provenance can be verified, and the integrity and uniqueness of the
asset is ensured.

\end{itemize}

Importantly, not all token-based payment systems use smart contracts or global
state transitions to facilitate transactions.  Also, depending upon how the
ledger is used, it might not be appropriate to measure ledger performance in
terms of number of transactions per unit time.  While such a metric might be
appropriate if the ledger establishes consensus about each transaction,
approaches that involve oblivious tracking or other mechanisms for ``rolling
up'' transactions, for example, those that aggregate a potentially large set of
transactions into a single hash value that subsequently becomes the subject of
consensus, can potentially achieve high throughput at scale without a
particularly fast core consensus engine.

\section{Using tokens}

Unlike balances, which represent the status of a relationship between an
account-holder and a fiduciary, the value of a token is intrinsic.  As digital
assets, tokens can be possessed and controlled in a manner akin to physical
assets, and there has been a trend toward increased recognition of the legal
status of digital assets, as evidenced by legislation recently introduced in
some jurisdictions, such as the Electronic Trade Documents Act (2023) in the
UK~\cite{etda2023}.

There are various different paradigms for how an owner can access and use
tokens.  Consider the difference between \textit{DLT accounts}, \textit{DLT
addresses}, and \textit{wallets} as defined in ISO 22739:

\begin{itemize}

\item\cz{DLT account:} ``representation of an entity participating in a
transaction in a DLT system''~\cite{iso22739}

\item\cz{DLT address:} ``data element designating the originating source or
destination of a transaction''~\cite{iso22739}

\item\cz{wallet:} ``application or mechanism used to generate, manage, store or
use private keys and public keys or other digital assets''~\cite{iso22739}

\end{itemize}

Some, but not all, token-based payment systems that use DLT make use of DLT
accounts or DLT addresses for the purpose of transactions, with implications
for control, possession, and privacy.

The difference between the locus of control of tokens contrasts with the
distinction between assets that are held on an owner’s device and assets that
are held by a third party.  The distinction is related to the difference
between control and possession:

\begin{itemize}

\item\cz{control:} transferable property of a relationship between an actor and
an asset wherein the actor and no other actor has the means to specify
legitimate changes to the asset

\item\cz{possession:} transferable property of a relationship between an actor
and an asset wherein the actor and no other actor can effect changes to the
asset

\end{itemize}

\begin{figure}[ht]
\begin{center}
\hspace{-0.8em}\scalebox{1.2}{
\begin{tikzpicture}[>=latex, node distance=3cm, font={\sf \small}, auto]\ts
\tikzset{>={Latex[width=4mm,length=4mm]}}
\node (box1) at (-2,2) [box, text width=10.7em, text height=10.7em] {};
\node (box2) at (2,2) [box, text width=10.7em, text height=10.7em] {};
\node (box3) at (-2,-2) [box, text width=10.7em, text height=10.7em] {};
\node (box4) at (2,-2) [box, text width=10.7em, text height=10.7em] {};
\node (n1) at (-2,4.5) {held by owner};
\node (n2) at (2,4.5) {held by third party};
\node[align=center] (t1) at (-2,2) {
    wallet with a\\
    hardware root of trust\\
    that enforces rules\\
    on behalf of an\\
    authority or issuer
};
\node[align=center] (t2) at (2,2) {
    online services that\\
    store digital assets\\
    and conduct transactions\\
    on behalf of the owner
};
\node[align=center] (t2) at (-2,-2) {
    hardware or software\\
    wallet accountable\\
    solely to the owner
};
\node[align=center] (t2) at (2,-2) {
    online services that\\
    store digital assets\\
    but for which the\\
    owner is responsible\\
    for their management\\
    and any transactions
};
\node[rotate=90] (n3) at (-4.5,2) {third party in control};
\node[rotate=90] (n4) at (-4.5,-2) {owner in control};
\end{tikzpicture}}
\end{center}

\caption{\cz{A characterisation of possession and control of digital assets.}}

\label{f:control-possession}
\end{figure}

Control differs from possession in that possession determines who can access an
asset, whereas control determines who can use an asset.  It is possible to have
possession without control, or control without possession.
Figure~\ref{f:control-possession} illustrates how wallets and online services
can be used to achieve different combinations of possession and control for
owners of digital assets:

\begin{itemize}

\item\cz{neither control nor possession}, wherein an authority or service
provider holds and transacts assets on behalf of a user and ultimately
determines whether and how an asset is transacted;

\item\cz{control without possession}, wherein a third party determines whether
an owner can access the assets (for example, if the assets are stored on media
that the owner cannot access direct) but only the owner can determine whether
and how to transact (for example, if transactions require a private key that
only the owner possesses);

\item\cz{possession without control}, wherein the owner determines who can
access the assets (for example, if they are stored exclusively on a device held
by the owner) but a third party must authorise or approve transactions (for
example, if the system requires the device to enforce rules specified by a
third party, via certified hardware with a hardware root of trust); and

\item\cz{both possession and control}, wherein the owner determines the media
on which the assets are stored (for example, an open source hardware or
software wallet) and has exclusive access to the data (for example,
cryptographic keys) required to authorise transactions.

\end{itemize}

In most cases in which a third party possesses the assets on behalf of an
owner, the third party acts as fiduciary for an account for which the owner is
an account-holder.  The account might or might not be a DLT account.

\section*{Acknowledgements}

We thank Professor Tomaso Aste for his continued support for our work on
digital currencies and digital payment systems.  We also acknowledge the Future
of Money Initiative at University College London and the Systemic Risk Centre
at the London School of Economics, and we acknowledge EPSRC and the PETRAS
Research Centre EP/S035362/1 for the FIRE Project.

\end{document}